\documentclass[a4paper]{article}

\usepackage{INTERSPEECH2021}
\usepackage{subfig}
\usepackage{xcolor}
\usepackage{multirow}
\usepackage[none]{hyphenat}

\title{Towards Intelligibility-Oriented Audio-Visual Speech Enhancement}
\name{Tassadaq Hussain$^1$, Mandar Gogate$^1$, Kia Dashtipour, Amir Hussain}

\address{Edinburgh Napier University}
\email{\{t.hussain,m.gogate,k.dashtipour,a.hussain\}@napier.ac.uk}
\begin{document}
\maketitle
\begin{abstract}
Existing deep learning (DL) based speech enhancement approaches are generally optimised to minimise the distance between clean and enhanced speech features. These often result in improved speech quality however they suffer from a lack of generalisation and may not deliver the required speech intelligibility in real noisy situations. In an attempt to address these challenges, researchers have explored intelligibility-oriented (I-O) loss functions and integration of audio-visual (AV) information for more robust speech enhancement (SE). In this paper, we introduce DL based I-O SE algorithms exploiting AV information, which is a novel and previously unexplored research direction. Specifically, we present a fully convolutional AV SE model that uses a modified short-time objective intelligibility (STOI) metric as a training cost function. To the best of our knowledge, this is the first work that exploits the integration of AV modalities with an I-O based loss function for SE. Comparative experimental results demonstrate that our proposed I-O AV SE framework outperforms audio-only (AO) and AV models trained with conventional distance-based loss functions, in terms of standard objective evaluation measures when dealing with unseen speakers and noises. 
\footnote{Equal contribution.}

% Different from deep learning (DL)-based approaches which are gradient-based neural architectures and are optimized using mean squared error (MSE), intelligibility-metric based DL approaches have proposed recently and shown to be effective in intelligibility enhancement. In this paper, we investigate the efficacy of DL-based intelligibility-oriented (I-O) speech enhancement (SE) algorithms using auditory and visual information which is a novel and previously unexplored direction. Therefore, we present a fully convolutional network (FCN)-based SE framework that uses short-time objective intelligibility (STOI) metric as a cost function to jointly optimize the audio-visual (AV)-based FCN model. To the best of our knowledge, this is the first attempt that combines AV modalities and optimize STOI metric using a FCN-based model. The experimental results demonstrate that the proposed I-O AV SE framework outperforms existing MSE- and STOI-based AO  and MSE-based AV SE frameworks with a reasonable margin in terms of standard objective evaluation measures when dealing with non-stationary and unseen environments. 

\end{abstract}

\section{Introduction}

The main goal of a speech enhancement (SE) system is to improve the quality and intelligibility of speech in real world environments where speech is often distorted by multiple-competing additive or convolutive noises. In the literature, extensive research has been carried out to develop SE methods for speech coding \cite{li2011two, li2011comparative, lim1979enhancement}, assistive hearing devices \cite{chern2017smartphone} \cite{wang2017deep} and automatic speech recognition (ASR) \cite{li2019multi} \cite{wang2020cross}. In recent years, machine-learning-based SE approaches have received great attention due to their ability to outperform state-of-the-art SE models. These approaches generally use machine-learning based mapping functions to reconstruct the clean speech from noisy input. Notable machine-learning-based SE approaches include sparse coding \cite{he2015spectrum}, robust PCA (RPCA) \cite{sun2014noise}, and non-negative matrix factorization (NMF) \cite{fan2014speech} \cite{smaragdis2014static}.

Recently, deep learning (DL) based models have been exploited in the SE field and yielded enhanced performance. For example, a deep denoising autoencoder (DDAE) framework has demonstrated promising SE performance compared to traditional methods \cite{lu2013speech}. Subsequently, a deep neural network (DNN) was adopted to handle a wide range of additive noises for the SE task \cite{xu2014regression}. In addition to standard feed-forward neural networks, different structures of convolutional neural networks (CNNs) have been employed in an attempt to improve the generalisation performance for SE. In \cite{pandey2019tcnn}, an audio-only (AO) CNN was trained in an encoder-decoder style with an additional temporal convolutional module to provide real-time SE. In \cite{fu2017raw}, a fully convolutional neural network (FCN) was exploited to effectively recover the enhanced speech waveform for AO SE in an end-to-end manner. Different from traditional DL-based approaches, authors in \cite{fu2018end} adopted a novel strategy and trained FCN using an objective evaluation-based cost function for enhanced speech perception. Research has shown that the visual modality carries important information (such as lip motions and mouth articulations) that can help discriminate similar speech sounds in noisy conditions. Recent examples on the use of multimodal approaches to address speech related issues by leveraging AV information to improve performance, include DL based AV SE systems, which have shown significant improvement in noise reduction. \cite{michelsanti2021overview, gogate2020cochleanet, wang2020robust, gogate2020deep, adeel2020contextual, gogate2020visual, michelsanti2021overview, gogate2018dnn}.

\begin{figure}[!t]
    \centering
    \includegraphics[height = 1.2in, width=0.5\textwidth]{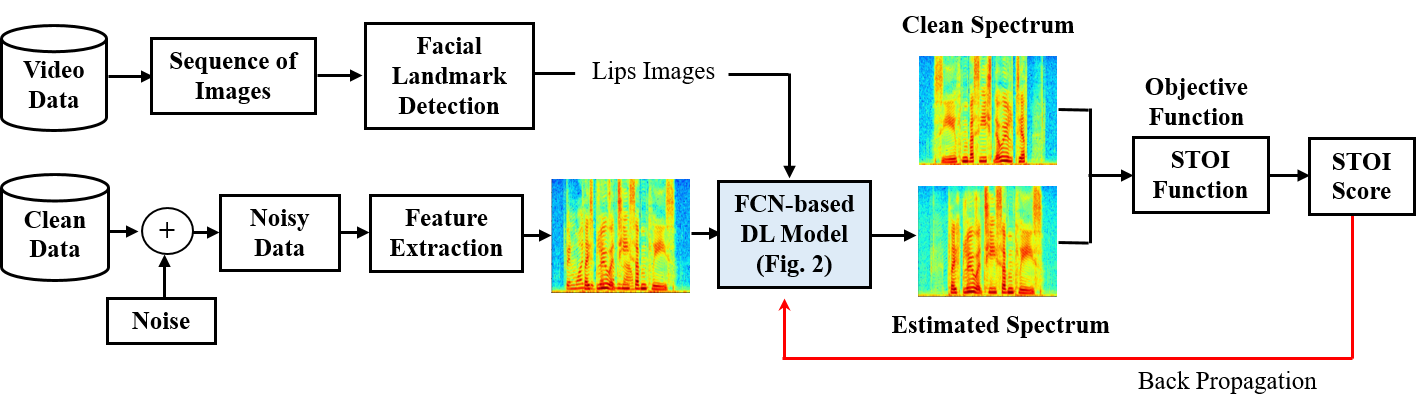}
    \caption{Block diagram of our proposed STOI-based audio-visual SE model}
    \label{fig:block}
\end{figure}

\begin{figure*}[!t]
    \centering
    \includegraphics[height = 2.4in, width=0.75\textwidth]{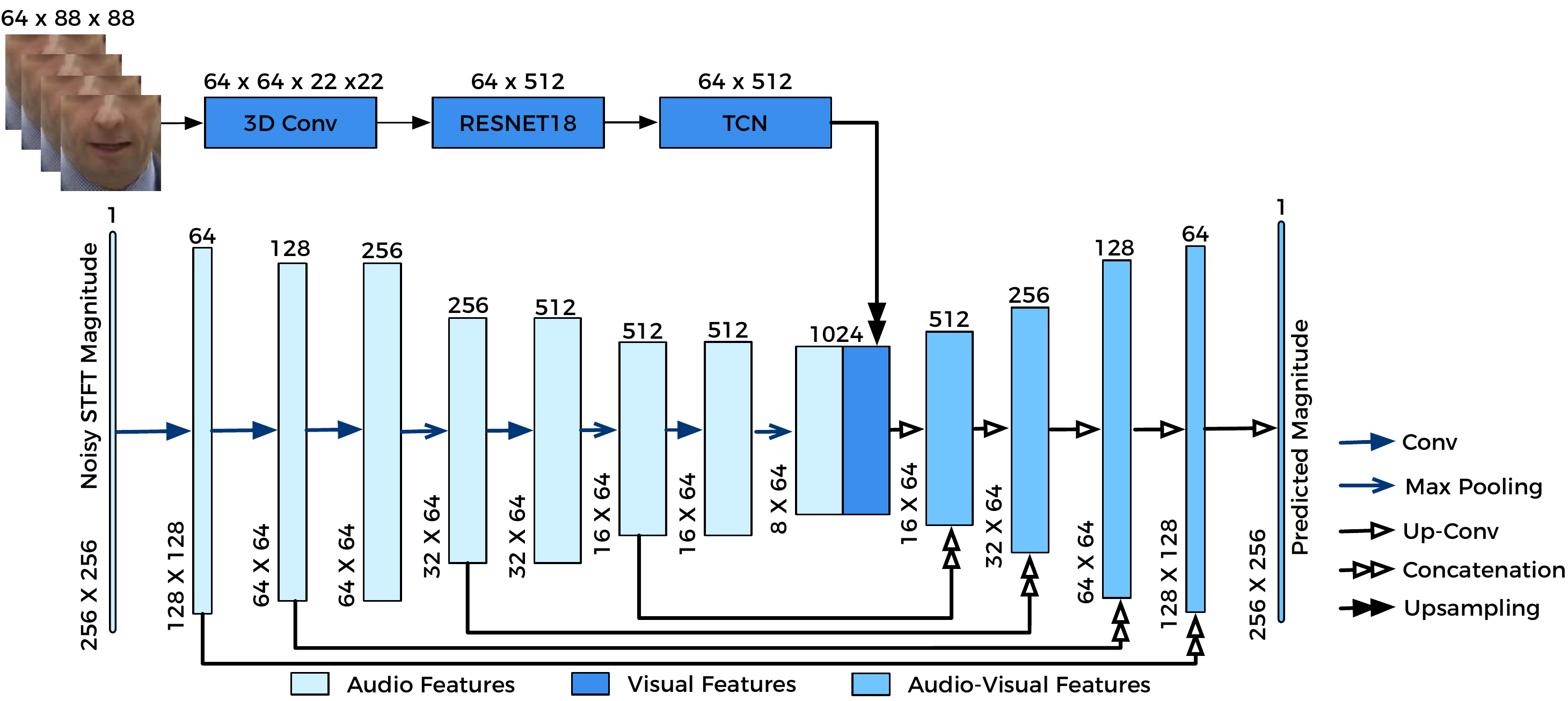}
    \caption{The FCN-based U-Net framework used to optimize the STOI-based audio-visual SE model}
    \label{fig:framework}
\end{figure*}

Despite the excellent performance achieved by DL based SE models, the parameters of such approaches are often optimized using distance-based loss functions including mean squared error (MSE) and mean absolute error (MAE). However, these  may not be optimal performance evaluation metrics for speech-related applications as they are not based on human auditory perception. In addition, we believe that the optimizing human perception-based evaluation metrics directly may lead to more optimal results corresponding to the target task. In the context of SE, researchers usually employ number of performance evaluation metrics that are inspired by human auditory perception. There are two widely used metrics, specifically, perceptual evaluation of speech quality (PESQ) \cite{rix2001perceptual} and short-time objective intelligibility (STOI) \cite{taal2011algorithm}, which are used to approximate subjective speech quality and intelligibility, respectively. Apart from conventional MSE/MAE-based DL approaches, a number of intelligibility-oriented (I-O) STOI-metric based DL approaches have also been proposed and shown to be effective for SE. For example, in \cite{fu2018end}, authors utilized the STOI measure as an objective function to optimize an AO fully-convolutional network (FCNN) model for SE. The results demonstrated that the STOI-based SE framework can perform significantly better than a conventional MSE-based SE system due to increased consistency between the training and evaluation target. In addition, authors in \cite{zezario2020stoi} proposed a DL-based speech intelligibility assessment model by combining a CNN and bidirectional long short-term memory (BLSTM) architecture with a multiplicative attention mechanism. More recently, authors in \cite{kolbaek2020loss} studied the influence of six different loss functions (including the STOI-based cost function) and evaluated them in a structured manner with end-to-end time-domain DL-based SE systems.

Motivated by the promising performance achieved by STOI-based SE systems, we further develop and extend conventional STOI-based AO SE approaches by incorporating visual information to jointly optimize the AV SE system in listening environments in which traditional methods can prove ineffective. The aim of this study is to investigate the effectiveness of STOI as an objective function to train DL-based AV SE architectures to overcome limitations of current frameworks. Unlike previously proposed STOI-based systems which process speech in an end-to-end manner to construct an utterance-based  (time-domain) SE system, we process the signal in a frame-wise manner in the frequency domain by focusing on magnitude spectra of noisy and clean speech utterances. We next use a FCN to learn the spectral mapping for AV input data and perform SE using a STOI-based cost function. This entails modifying conventional (classical and extended) STOI measures to account for signals in the frequency domain. In addition to formulating the modified STOI-based cost function, we comparatively evaluate the effectiveness of two conventional distance-based cost functions, namely MSE and MAE, to optimise AV SE system performance. All AV SE frameworks are trained and evaluated using a two-speaker synthetic mixture of the benchmark GRID corpus \cite{cooke2006audio}, at random Signal-to-Noise Ratios (SNR). Note that the task of suppressing speech interference is more challenging than suppressing non-speech noises. Experimental results show that our proposed AV framework optimized with a modified STOI-based cost function can achieve significant SE performance improvement over both  MSE and MAE-based AO and AV frameworks, as well as recently proposed STOI-based AO methods, under mismatched testing conditions, using a range of standardized objective measures: namely, the perceptual evaluation of speech quality (PESQ), STOI, scale-invariant signal-to-distortion ratio (SI-SDR) \cite{le2019sdr}, and Virtual Speech Quality Objective Listener (VISQOL) \cite{hines2015visqol}.

The remainder of the paper is organised as follows. Section 2 describes the methodology and our proposed STOI-based AV SE system. Section 3 presents the experimental setup including dataset description, AV feature extraction and comparative evaluation results. Finally, concluding remarks are presented in Section 4.

\begin{figure*}[!t]
        \centering
    \subfloat[\centering STOI vs Modified STOI]{{\includegraphics[width=0.4\textwidth]{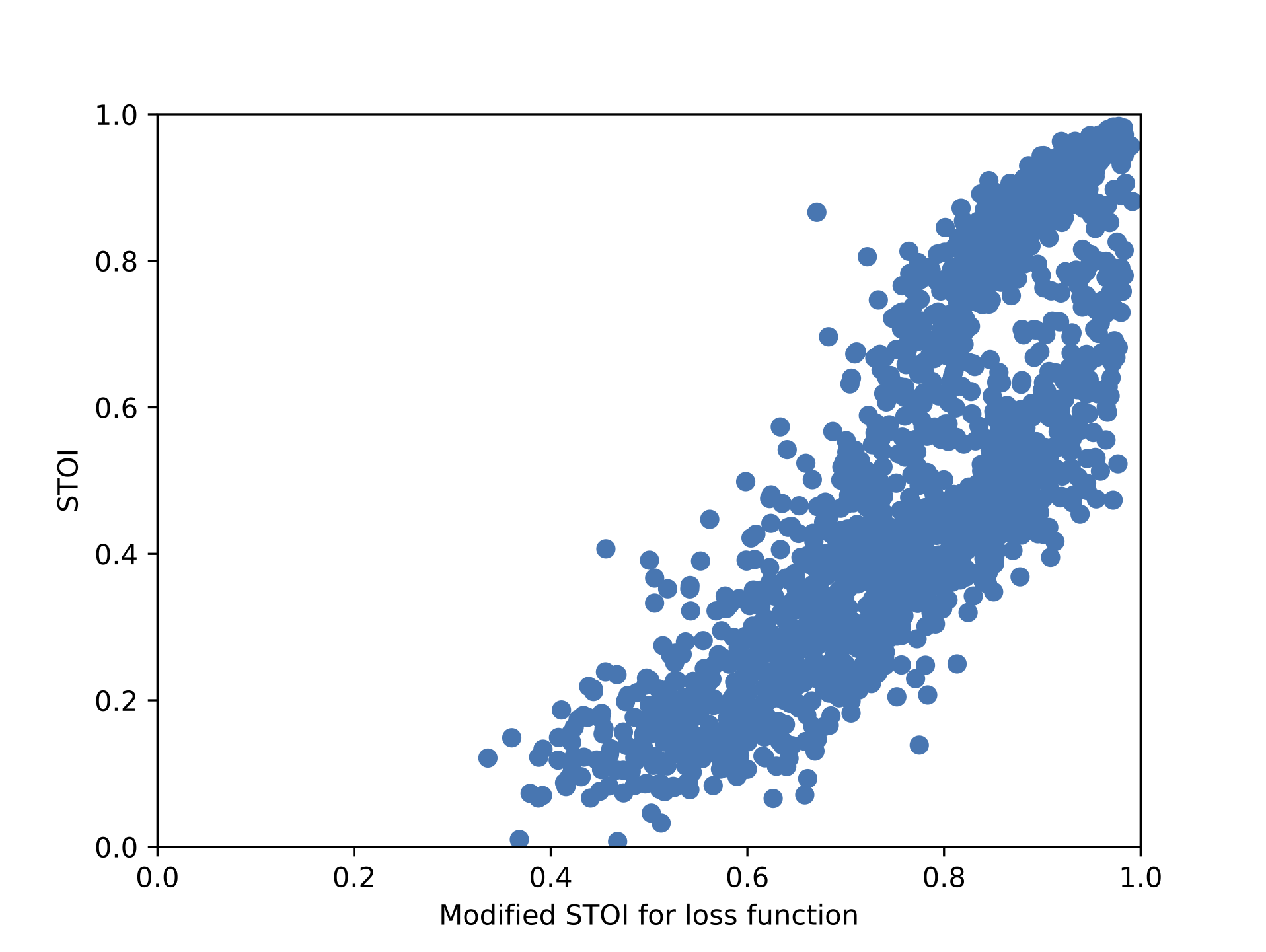} }}%
    \qquad
    \subfloat[\centering Extended STOI vs Modified Extended STOI]{{\includegraphics[width=0.4\textwidth]{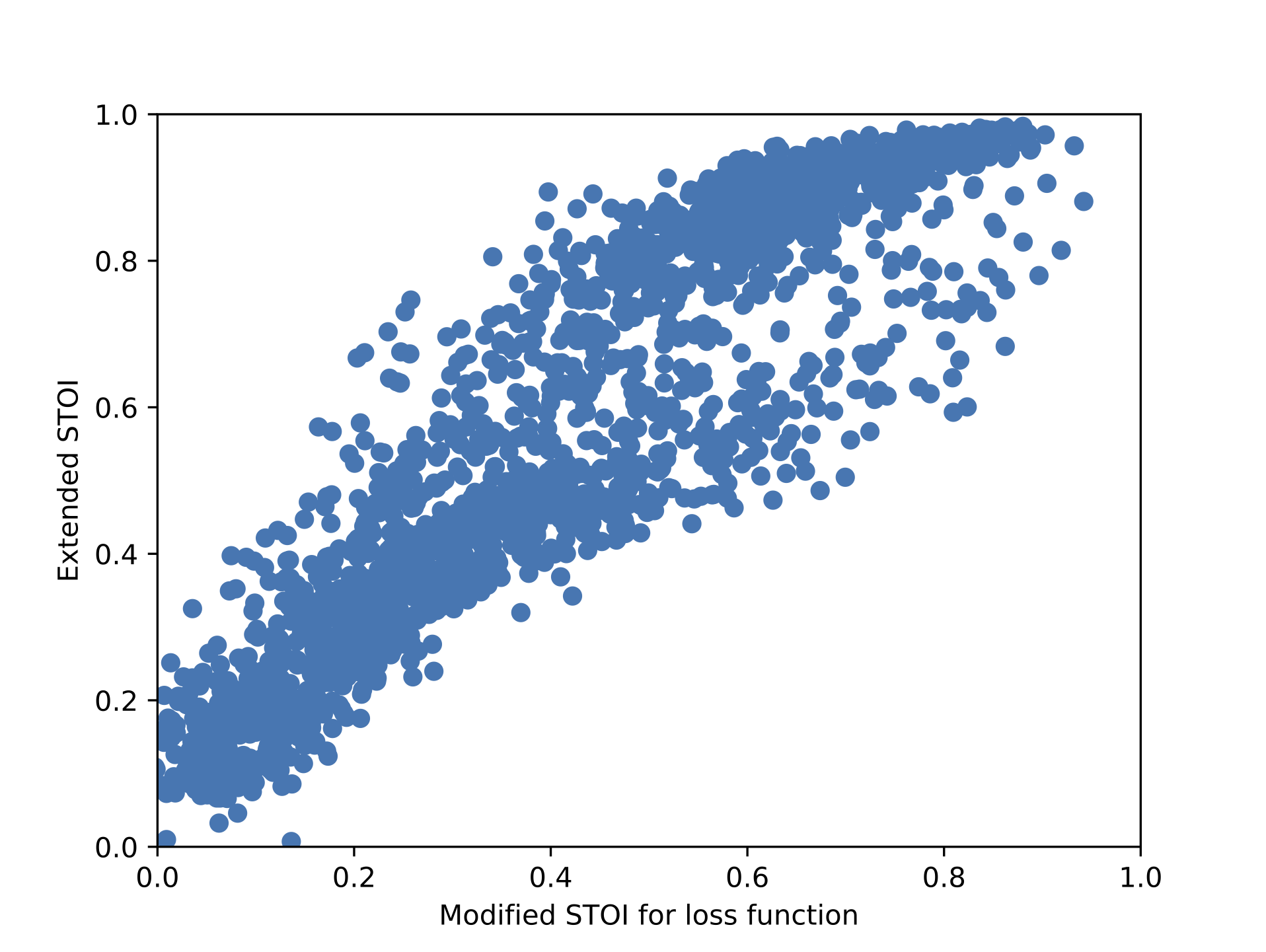} }}%
    \caption{Scatter Plot between (a) STOI vs Modified STOI and (b) Extended STOI vs Modified Extended STOI}%
    \label{fig:example}%
\end{figure*}

%\begin{figure}[h]
%    \centering
%    \includegraphics[height = 0.8in, width = \linewidth]{ASE.pdf}
%    \caption{Intelligibility-oriented Audio-only Speech Enhancement Framework.\label{fig:ASE}} 
%\end{figure}

%%In real noisy conditions, human listening performance is known to be dependent upon both visual senses also that are contextually combined with the aural by the brain’s multi-level integration strategies. The multimodal nature of speech is well established, with listeners known to unconsciously lip read to improve the intelligibility of speech in noise. Further, looking at a speaker makes speech more detectable in noise, i.e., as if audio cues are being visually enhanced. In this study, we first examine the case for using the visual information in A-only I-O SE to further enhance the intelligibility (listening) in listening environments in which traditional A-only I-O frameworks currently prove ineffective. Second, the study will address the issues associated with A-only I-E SE frameworks, such as the ability to assign meaning to sound in extremely noisy environments, which is a key challenge for current A-only I-O SE frameworks. 

\section{Methodology}

%For human to human communication, we care more about speech quality and intelligibility. Normally, we conduct listening and recognition tests to analyse the quality and intelligibility of the estimated speech signal. However, it is not feasible to conduct human listening tests for a large number of samples. As a result, two evaluation metrics, PESQ and STOI have been developed to assess speech quality and intelligibility, respectively. 
Conventional DL-based SE models are trained using MSE and MAE loss functions. Thus, we first consider the  most commonly used MSE loss function to train a FCN based AV SE model, which is implemented using the benchmark U-Net framework ~\cite{ronneberger2015u}. The MSE can be computed as follows:
 \begin{equation}
 \mathcal{L}_{MSE} = min (\frac{1}{M} \sum_{m=1}^{M} \|\hat{Y}_m - Y_m\|_2)
\end{equation} 
where $\textit{M}$ is the total number of speech frames, $\textit{\^Y}_M$ is the estimated magnitude spectrum, $\textit{Y}_M$ is the reference magnitude spectrum of the utterance, and $ \| \cdot \|_2$ denotes L2-normalization.

Recent studies have shown that the STOI metric is highly correlated to human perception and performs more optimally compared to MSE, suggesting it could be employed as an alternative loss function in speech-related applications \cite{fu2018end} \cite{kolbaek2020loss}. In this paper, we evaluate the effectiveness of a STOI based loss function to optimise the performance of a DL-based AV SE system. Specifically, we adopt a deep FCN-based U-Net architecture that takes the noisy magnitude spectrum as input and exploits a modified STOI loss function to optimally learn the spectral mapping and estimate the enhanced magnitude spectrum. Figure 1 shows the block diagram of our proposed I-O AV SE framework, and Figure 2 illustrates the FCN-based U-Net framework used to optimize our STOI-based AV SE model. Our key focus is to explore how incorporating visual information into an AO SE model and optimisation using a modified STOI loss function, impacts the quality and intelligibility of the enhanced speech signal as evaluated with a range of standardized objective measures

\subsection{Short-time Objective Intelligibility (STOI)}
Here, we develop and employ a modified version of a well-known STOI intelligibility measure as an objective function to train AV SE models. The STOI is an intrusive measure that requires both estimated speech and reference (clean) speech signals and ranges from 0 to 1, with 1 denoting the highest intelligibility of the speech signal. The STOI function takes the clean and estimated speech signals as input and computes the score in five steps: i) Removal of silent regions from clean and estimated speech signals, ii) Application of the short-time Fourier transform (STFT); iii) Estimation of the short-time envelope of clean and noisy speech using one-third octave-band analysis of the STFT frames; iv) Normalization and clipping to compensate for global level differences and stabilisation of the STOI evaluation; and v) Intelligibility measure computation: the correlation coefficient between the two spectral envelopes is estimated using the equations below.

 \begin{equation}
 \textit{d}_{i,j} = \frac {(y_{i,j} - \mu_{y_{i,j}})^\intercal  (\hat{y}_{i,j} - \mu_{y_{i,j}})^\intercal } {\|y_{i,j} - \mu_{y_{i,j}}\|_2  \|\hat{y}_{i,j} - \mu_{\hat{y}_{i,j}}\|_2}
\end{equation} 
 where $\textit{y}$ and $\textit{\^y}$ are the short-time spectral envelope of the reference clean and estimated speech signals, $\mu_{y_{i,j}}$ and $\mu_{\hat{y}_{i,j}}$ are the corresponding sample mean vectors, and $\| \cdot \|_2$ represents the L2-normalization. The final STOI is the average of the intelligibility measure over all bands and frames.
 
 \begin{equation}
 \textit{d}_{STOI} = \frac {1} {I(M - N + 1)} \sum_{i=1}^{I} \sum_{j=1}^{J} \textit{d}_{i,j}
\end{equation} 
 where $I$ = 15 is the number of one-third octave band and $M - N + 1$ is the total number of short-time temporal envelope vector. For a more detailed setting of each step, please refer to \cite{stoi}. The computation of STOI is differentiable, thus, it can be used as the objective function directly to optimize the AV SE model.
 
 \begin{equation}
 %\mathcal{L}_{STOI} = - \frac {1} {K} \sum_{m=1}^{K} \textit{d}_{STOI}(\^Y_m, Y_m) + \textit{d}_{STOI}(\^V_m, V_m)
  \mathcal{L}_{STOI} = - \frac {1} {M} \sum_{m=1}^{M} \textit{d}_{STOI}(\hat{Y}_m, Y_m)
\end{equation} 
where $\textit{d}_{STOI}(\hat{Y}_m, Y_m)$ measures the STOI score between the estimated and clean magnitude spectra of audio utterances. Unlike the MSE, where the goal is to reduce the distance, we want to maximize the STOI score to enhance speech intelligibility.
% {\color{blue}where $\textit{Z}_m = [\hat{Y}_m, V_m]$ represents the concatenation of audio and visual features and $\textit{d}_{STOI}(\hat{Z}_m, Z_m)$ jointly optimize the AV framework by measuring the STOI score between the estimated and clean features concatenated magnitude spectra of audio utterance and $\textit{d}_{STOI}(\hat{V}_m, V_m)$ computes the STOI score between the corresponding reconstructed and clean lip image, respectively. Unlike MSE, where the goal is to reduce the distance, we want to maximize the STOI score for better speech intelligibility.}

\subsection{Proposed Audio-Visual SE Framework}

This section presents the DL models used for our I-O AV SE framework as depicted in Fig.~\ref{fig:framework}. Specifically the network architecture for required AV feature extraction, fusion and speech resynthesis pipeline is outlined below. 

\subsubsection{Audio feature extraction}
The audio feature extraction stage utilises a U-net~\cite{ronneberger2015u} style network consisting of an encoder and decoder block modified for AV SE. The input to the network is the magnitude of noisy speech Short-Time Fourier Transform (STFT) of dimension $F \times T$ where $F$ and $T$ are frequency and time dimension of the spectrogram. The input is fed to two convolutional layers with filter size of 4 and stride of 2 to downsample the time-frequency dimension until the time dimension is equal to 64. The downsampled features are passed through three convolutional blocks each consisting of two convolutional layers with filter size of 3 and stride of 1, followed by a frequency pooling layer that reduces the frequency dimension by 2. Note that the spatial dimension is preserved during the processing of convolutional blocks. 

\subsubsection{Visual feature extraction}
The visual feature extraction stage of the pipeline comprises a 3D convolutional layer with filter size of $5\times7\times7$ and stride of $1\times2\times2$, followed by RESNET-18 \cite{he2016deep}. The residual network features are then fed to a temporal convolutional network (TCN) as described in \cite{martinez2020lipreading}. The input to the network is a time-series of lip cropped images of size $N \times 88 \times 88$, where N is the number of frames. The visual feature network outputs a 512-D vector for each lip image. The visual features are upsampled to match the audio feature sampling rate.

\subsubsection{Multimodal fusion}
The upsampled visual features and audio features are concatenated and fed to a U-net decoder as shown in Fig.~\ref{fig:framework}. The decoder consists of 3 up convolutional blocks each consisting of two upsampling layers that upsample the time dimension by 2, followed by convolutional layers with a filter size of 3 and stride of 1. The AV features are then fed to two transposed convolutional layers with filter size of 4 and stride of 2 to upsample the time-frequency dimension, until the time-frequency dimension is equal to the input. Next we use a sigmoid layer to map the output in the range of 0 to 1. The predicted mask is then multiplied with the input spectrogram to generate the masked spectrogram as output.  

\subsubsection{Speech Resynthesis}
The proposed model estimates the clean spectrogram when the noisy spectrogram and cropped lip images are fed as input. The estimated magnitude is combined with the noisy phase to generate enhanced speech using an inverse STFT. 

\section{Experiments and Results}

\begin{table*}[t]
\normalsize
%\captionsetup{justification=centering}
\caption{\scshape Performance comparison of Audio-only and Audio-Visual DNN models using standardised objective evaluation metrics under speaker-independent conditions.}
\centering
\begin{tabular}{c|c|ccccccc|c}
\hline
\hline
\multirow{2}{*}{\textbf{Framework}}  & \multirow{2.2}{*}{\textbf{Loss Function}} & \multicolumn{7}{c|}{\textbf{Objective Evaluation Metrics}} & \multirow{2.2}{*}{\textbf{Avg.}} \\ \cline{3-9}

  &  & \multicolumn{1}{c}{PESQ} & \multicolumn{1}{c}{STOI} & \multicolumn{1}{c}{SI-SDR} & \multicolumn{1}{c}{CSIG} & \multicolumn{1}{c}{CBAK} & \multicolumn{1}{c}{COVL} & \multicolumn{1}{c|}{VISQOL}  \\ \cline{2-9}
 
\hline
\hline
\textbf{Noisy} & -- & 2.414 
& 0.828 & 8.067 & 3.159 & 2.441 & 2.373 & 3.213 & 3.214 \\ \hline
\ & MSE & 2.712 & 0.851 & 10.441 &  3.515 & 2.815 & 2.783 & 3.268 & 3.769\\
\textbf{Audio-only} & MAE & 2.789 & 0.852 & 10.794 & 3.705 & 3.004 & 3.017 & 3.301 & 3.923 \\
 & STOI & 3.005 & 0.884 & 11.246 & 3.852 & 2.781 & 3.132 &  3.376 & 4.039 \\  \hline
 \ & MSE & 2.724 & 0.857 & 10.640 & 3.644  & 2.847 & 2.861 & 3.284 & 3.836  \\
 \textbf{Audio-Visual} & MAE & 3.008 & 0.887 & 11.753 & \textbf{3.991} & \textbf{3.143} & \textbf{3.274}  & 3.403 &  4.208\\ 
  \ & STOI & \textbf{3.206} &\textbf{0.914} & \textbf{12.403} & 3.863 & 2.844 & 3.184 & \textbf{3.478} &  \textbf{4.270}\\
  \hline
\textbf{IRM} & -- & 3.432 & 0.872 & 7.383 & 4.683 & 2.389 & 3.902 & 3.501 & 3.737 \\
\hline
\hline
\end{tabular}
%}
     \end{table*}
     
\subsection{Experimental Setup}

For initial testing, we trained our proposed I-O AV SE model using a small vocabulary AV corpus to assess how the STOI loss function affects the overall SE performance. Specifically, the performance of  the framework was evaluated using the benchmark GRID corpus \cite{cooke2006audio}. The dataset contained AV recordings of clean utterances from 34 male and female speakers, each with 1000 utterances lasting around three seconds. The AV utterances were recorded in a quiet room with sufficient background lighting, with the speaker filmed facing the camera. The visual data was captured at a frame rate of 25 frames per second (fps) while the audio data was recorded at 48 kHz which was then resampled to 16 kHz. We randomly selected 23 speakers for the training set and 4 speakers each for the, validation and test sets. The split ensured speaker independence criteria i.e. there was no overlap of speakers between the training, validation and test sets. The test and validation set comprised 2 male and 2 female speakers. The clean utterances were mixed with randomly selected clean speech utterances from the respective sets, at randomly selected SNRs ranging from [0 to 20] dB with 1 dB increments. In total, training, validation and testing had 46000, 4000 and 4000 utterances respectively. To improve generalisation during training a single clean utterance was mixed with two different randomly selected interference. In order to objectively measure the quality of denoised speech, a number of state-of-the-art evaluation metrics were used, including PESQ, STOI, SI-SDR and VISQOL. Furthermore, as proposed in \cite{hu2007evaluation}, three additional measures were used to: compute the signal distortion of the speech signal (termed CSIG), predict the background noise in the estimated signal (termed CBAK), and predict the overall quality of the estimated speech (termed COVL)
% In addition to cost function design, we used a STOI measure as an evaluation metric to assess the speech intelligibility performance of our proposed framework. Apart from the primary objective evaluation metrics, PESQ, STOI, SI-SDR, and VISQOL, we employed three additional measures to; compute the signal distortion of the speech signal (termed CSIG), predict background noise in the estimated signal (termed CBAK), and predict overall quality of the estimated speech (termed COVL), as mentioned in \cite{hu2007evaluation}.

%%%%%%%%%%%%%%%%%%%%%%%%%%%%%%%%%%%%%%%%%%%%%%%%%%%%%%%%%%%%%%%%%%%%%%%%%%

\subsection{Audio and Visual Features}
We used the STFT with a frame length of 25ms and a frameshift of 10ms to process the audio speech signals. For the visual features, we converted each video into a sequence of images at a frame rate of 25 fps. The lip region of size 88 x 88 was extracted using the Dlib library and extracted lip regions were converted to greyscale. 

\subsection{STOI vs Modified STOI}
Unlike the original (classical and extended) STOI measures, which initially down sample speech signals to 10kHz, carry out silent frame removal, and then apply STFT, we formulated a modified version of STOI (termed modified STOI) to account for 16kHz signals in the frequency domain while ignoring downsampling and silent frame removal steps. To examine the behaviour of the modified STOI, we plotted correlations between the modified STOI and original STOI (classical and extended) scores as shown in Fig. 3. Specifically, Fig. 3 (a) and (b) present scatter plots for STOI (modified STOI vs original STOI) and (modified extended STOI vs extended STOI) scores, respectively. From the figures, we can note that the modified STOI scores are strongly correlated with the original and extended STOI scores, demonstrating that our modified extended STOI correlated well with the extended STOI and can be directly used as a loss function to train and optimize the DL models for AV SE.

\subsection{Objective Evaluation}
We investigated the impact of our modified STOI loss function on the performance of the frequency-domain AV SE system in terms of PESQ, STOI, SI-SDR, CSIG, CBAK, COVL, and VISQOL. We used the same setup to train three AO and AV SE frameworks utilizing three different loss functions, namely $\mathcal{L}_{MSE}$, $\mathcal{L}_{MAE}$, and $\mathcal{L}_{STOI}$, and evaluated their performance using the GRID dataset (see Sec. 3.1).   

Table I shows the performance comparison of AO and AV SE frameworks trained using three different loss functions. It can be seen from Table I that, both AO and AV SE frameworks, when trained with different loss functions, enhanced the original noisy speech utterances with a reasonable margin in terms of all performance measures. In short, both frameworks optimized using three loss functions proved to be effective for SE. We note that in contrast to AO SE frameworks trained with $\mathcal{L}_{MSE}$ and $\mathcal{L}_{MAE}$ functions, the AO framework trained with the $\mathcal{L}_{STOI}$ loss function demonstrated better performance in terms of PESQ, STOI, SI-SDR, CSIG, CBAK, COVL, and VISQOL scores, respectively. 

\begin{figure*}[!t]
    \centering
    \includegraphics[width=\textwidth]{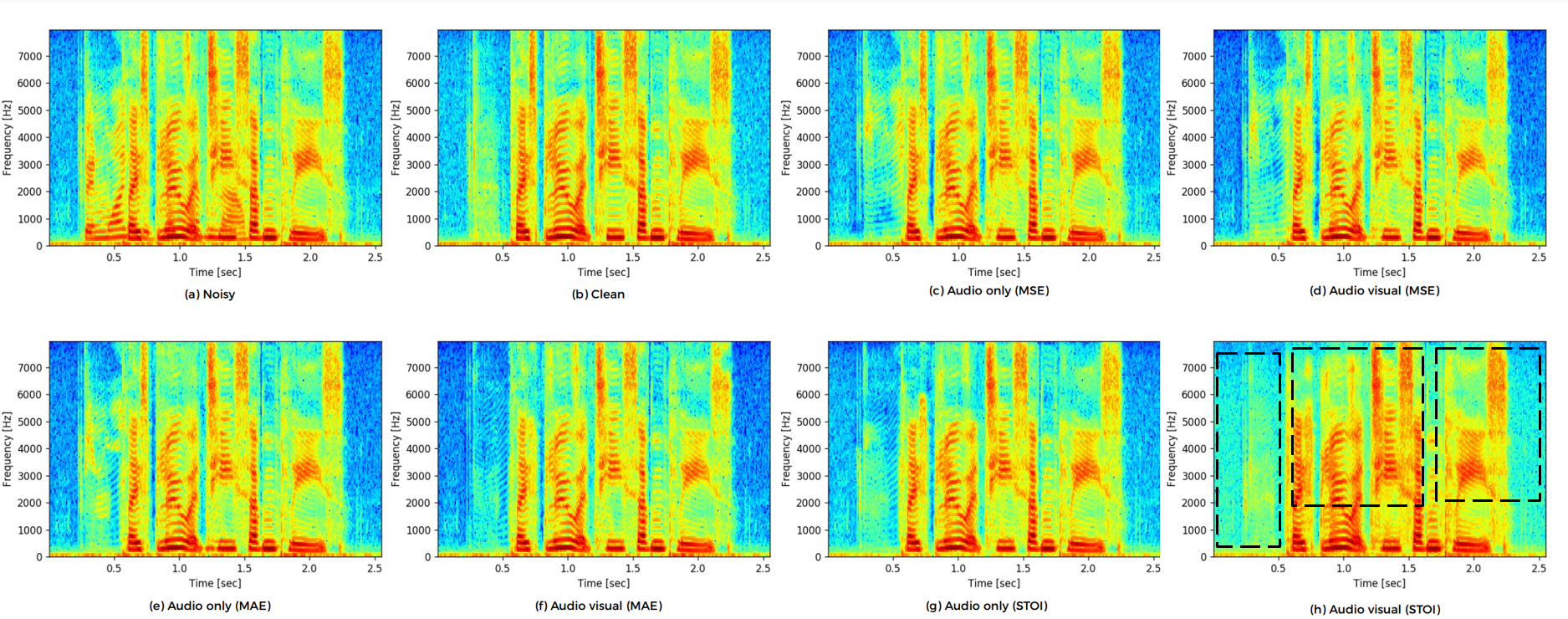}
    \caption{Spectrogram of a randomly selected utterance from the test set. It can be seen that our modified STOI based AV SE model recovered more speech regions than conventional AO and AV SE models} 
\end{figure*}

Further, we note that despite the excellent performance of AO SE systems optimised using three loss functions, it can be seen from Table I that incorporating visual information into the AO systems and training with $\mathcal{L}_{MSE}$, $\mathcal{L}_{MAE}$, and $\mathcal{L}_{STOI}$ functions, further improves not only primary objective evaluation metrics like the PESQ, STOI, SI-SDR, and VISQOL, but also other metrics like the CSIG, CBAK, and COVL. In particular, we observe that our proposed AV SE framework, when trained with $\mathcal{L}_{STOI}$, improves the performance significantly in terms of PESQ, STOI, SI-SDR, and VISQOL. However, the framework under-performs when compared with $\mathcal{L}_{MAE}$ loss for the CSIG, CBAK, and COVL measures. It is to be noted that the performance of the AO  SE framework optimised using the STOI loss function is similar to the AV SE  framework trained using the MSE loss function. This shows that the performance improvement achieved using a distance-based metric and AV information is similar to one achieved using an I-O loss function.

Finally, we plot the spectrograms of randomly selected clean and noisy speech signals and compare them with enhanced speech signals estimated by AO  and AV SE frameworks using MAE, MSE and STOI loss functions. Figures 4(a) and (b) display the spectrogram of a noisy test utterance contaminated by a female speaker's speech at 3 dB SNR and corresponding clean speech signal. Figures 4(c) and (d) show the spectrograms of the enhanced speech signal for AO  and AV SE frameworks optimised using $\mathcal{L}_{MSE}$. Similarly, Fig. 4(e), (f), (g), and (h) present the spectrogram of the enhanced speech signal for the two frameworks optimised using $\mathcal{L}_{MAE}$ and $\mathcal{L}_{STOI}$. It can be seen that despite the excellent performance achieved by MSE and MAE-based AO and AV SE frameworks in terms of objective evaluation measures and noise suppression capabilities (evident from Fig. 4(c)--4(f)), these frameworks were unable to capture some middle and high-frequency regions when compared with STOI-based frameworks (Fig. 4(g)-(h)). Specifically, our proposed STOI-based AV SE framework (Fig. 4(h)) restored more (low-mid-high frequency region) speech components compared to MSE- and MAE-based AO and AV SE frameworks (as illustrated with dashed boxes in Fig. 4).

\section{Conclusion}
In this paper, we proposed a novel I-O AV SE paradigm to enhance the performance of conventional AO SE systems by exploiting an intelligibility-based evaluation metric as an alternative cost function.  Specifically, we developed and utilised a modified version of the conventional STOI loss function to train AV SE models, that can effectively account for signals in the frequency domain as opposed to conventional I-O frameworks that require down sampling signals in the time-domain. Comparative experimental findings show that incorporating visual information as part of a modified STOI-based AV DL framework can estimate the output signal with enhanced speech quality and intelligibility. In summary, we found that an I-O based loss function achieves good general performance and produces better results for a variety of SE evaluation metrics, implying that the modified STOI is a promising choice to optimize frequency-domain AV SE applications. Ongoing work is aimed at evaluating our I-O AV SE system with more challenging real-world AV corpora and subjective listening tests for speech and hearing-aid applications.  
%From Amir (mention/prioritise these ideas in our next week's COG-MHEAR postdoc update Workshop, as next extended version of this paper for a journal): May be better to remove from this brief paper so dont give away these good ideas! In future, we intend to apply canonical correlation analysis (CCA) as an alternative cost function by replacing the standard correlation in STOI with CCA. In addition, we will conduct a comparative analysis of qualitative/intelligibility metrics using novel CCA-based formulations, and investigate their potential usage in AV speech in noise tests. 

\section{Acknowledgements}
This work is supported by the UK Engineering and Physical Sciences Research Council (EPSRC) programme grant: COG-MHEAR (Grant reference EP/T021063/1).

\bibliographystyle{IEEEtran}
\bibliography{mybib}
\end{document}